\documentclass[11pt,twoside]{article}
\usepackage{asp2010}

\resetcounters
\aspvolume{455}
\aspcpryear{2012}
\aspvoltitle{4$^{th}$ {\em Hinode} Science Meeting: Unsolved 
Problems and Recent Insights}
\aspvolauthor{L.~R.~Bellot Rubio, F.~Reale, and M.~Carlsson, eds.}

\begin{document}

\title{{\em Hinode}/EIS Line Profile Asymmetries and Their Relationship 
\\ with the Distribution of SDO/AIA Propagating Coronal Disturbance Velocities}
\author{Marybeth~Sechler,$^1$ Scott~W.~McIntosh,$^1$ Hui~Tian,$^1$ and Bart 
De Pontieu$^2$}
\affil{$^1$High Altitude Observatory, National Center for Atmospheric Research, 
P.O.~Box~3000, Boulder CO 80307, USA}
\affil{$^2$Lockheed Martin Solar and Astrophysics Laboratory, Palo Alto, 
CA 94304, USA}

\begin{abstract}
Using joint observations from {\em Hinode}/EIS and the Atmospheric Imaging
Array (AIA) on the Solar Dynamics Observatory (SDO) we explore the
asymmetry of coronal EUV line profiles. We find that asymmetries exist
in all of the spectral lines studied, and not just the hottest lines
as has been recently reported in the literature. Those asymmetries
indicate that the velocities of the second emission component are
relatively consistent across temperature and consistent with the
apparent speed at which material is being inserted from the lower
atmosphere that is visible in the SDO/AIA images as propagating
coronal disturbances. Further, the observed asymmetries are of similar
magnitude (a few percent) and width (determined from the RB analysis)
across the temperature space sampled and in the small region
studied. Clearly, there are two components of emission in the
locations where the asymmetries are identified in the RB analysis,
their characteristics are consistent with those determined from the
SDO/AIA data. There is no evidence from our analysis that this second
component is broader than the main component of the line.
\end{abstract}

\section{Introduction}

Recently, strong upflows of between 50 and 150~km~s$^{-1}$ have been
observed over a broad range of temperatures spanning the chromosphere,
through the transition region, and into the corona
\citep[][]{2011Sci...331...55D}. These flows have been associated with
pronounced asymmetries in the blue wing of the spectral line profiles
observed and have been proposed as the mechanism by which the corona
is filled with hot plasma \citep[][]{2009ApJ...701L...1D,
2009ApJ...706L..80M}. However, the characteristic behavior of this
phenomena is still a topic of some debate
\citep[][among others]{2008ApJ...678L..67H, 2010ApJ...715.1012B,
2011ApJ...727...58W, 2010A&A...521A..51P}. While most of the
literature can agree that the observations contain more than one
emitting component that is not instrumental in origin, a product of
spectral blending, or the result of geometric projection effects, the
specific details of these components have not been sufficiently
consolidated, or explained. Indeed, many of the viewpoints are at
significant odds with each other. \citet{2009ApJ...701L...1D}, and
subsequent papers, consider the background dominant emission component
to be that of the corona which has low velocities compared to the
weak, superimposed second component which occurs at high velocities
and does not vary excessively with
temperature. \citet{2010A&A...521A..51P} also asserted that the
asymmetric profile is composed of two components. However, he
suggested that the bulk of the emission originates in a relatively
narrow ($\sim$50~km~s$^{-1}$) unshifted component that is superimposed on a
significantly broader (${>}$100~km~s$^{-1}$) ``pedestal'' of emission that is
shifted only slightly (10--20~km~s$^{-1}$) to the blue of the main
component. This is similar to the concept proposed in
\citet{2000A&A...360..761P, 2001A&A...374.1108P}. However, the
interpretation of the secondary component has now been shifted towards
spicule-like motions and Alfv\'{e}nic turbulence
\citep{2010A&A...521A..51P}. \citet{2010ApJ...715.1012B} found
outflows with much higher velocities for the second component of the
emission than \citet{2010A&A...521A..51P}, stating them to be as high
as 200~km~s$^{-1}$. They also reported
\citep[like][]{2011ApJ...727...58W} that the nature of the second
component had a temperature dependence, in which outflows were only
observed in hot lines (those formed above the formation temperature of
the Fe~XII 195\AA{}).

In essence we all have the same objective: determine the model of the
emission that is minimally consistent with the observed line profiles,
but just how do we achieve that? In this short paper we will take some
primitive steps in that direction---using a combination of high
signal-to-noise imaging observations from the Atmospheric Imaging
Assembly (AIA) on the {\em Solar Dynamics Observatory} (SDO) to study
the observed propagating coronal disturbances
\citep[][]{2011Sci...331...55D} in combination with the spectra
observed by {\em Hinode}/EIS \citep[][]{2007SoPh..243...19C} of the
same region. The target was NOAA Active Region (AR) 11106 that was
just to the south and east of disk center on September 16, 2010. In
the following section we provide some brief details of the sequence
and identify a small fan region for further study. 

Our study combines two different techniques; the first being the RB
analysis \citep[][]{2009ApJ...701L...1D} of the line profiles in that
region, and the second using an SDO/AIA image time series to track the
apparent motion of the propagating coronal disturbances (PCDs) rooted
at the base of the asymmetric regions \citep[see,
e.g.,][]{2009ApJ...706L..80M}. We find that asymmetries exist in all
clean spectral lines studied, and not just the hottest, though these
asymmetries can ``hide'' in certain spectral windows if the behavior
of the local continuum emission is not properly accounted for. Those
asymmetries indicate that the velocities of the second emission
component are relatively consistent across temperature and consistent
with the speed at which material is being inserted from the lower
atmosphere \citep[][]{2009ApJ...706L..80M}. Further, the asymmetries
are of similar magnitude (a few percent) and have an 1/e width
(determined from the RB analysis) that are consistent across the lines
observed and the region chosen for detailed analysis in this short
paper.

Clearly, the data analysis shows that more than one emission component
is present in some locations. Understanding the make-up of these line
profiles is critical to understanding how mass is cycled from the
lower atmosphere into the corona, before cooling and returning some
time later. The spectra that we detect encode all of these phases of
heating and cooling and, as such, represent the most we can possibly
learn about the coronal plasma with current instrumentation. The
observations indicate that a more sophisticated approach is needed to
analyze these spectral datasets, one which should also be used to
re-assess the archived data to identify the sources and signatures of
heated mass motion throughout the outer solar atmosphere.

\articlefigure[width=.96\textwidth]
{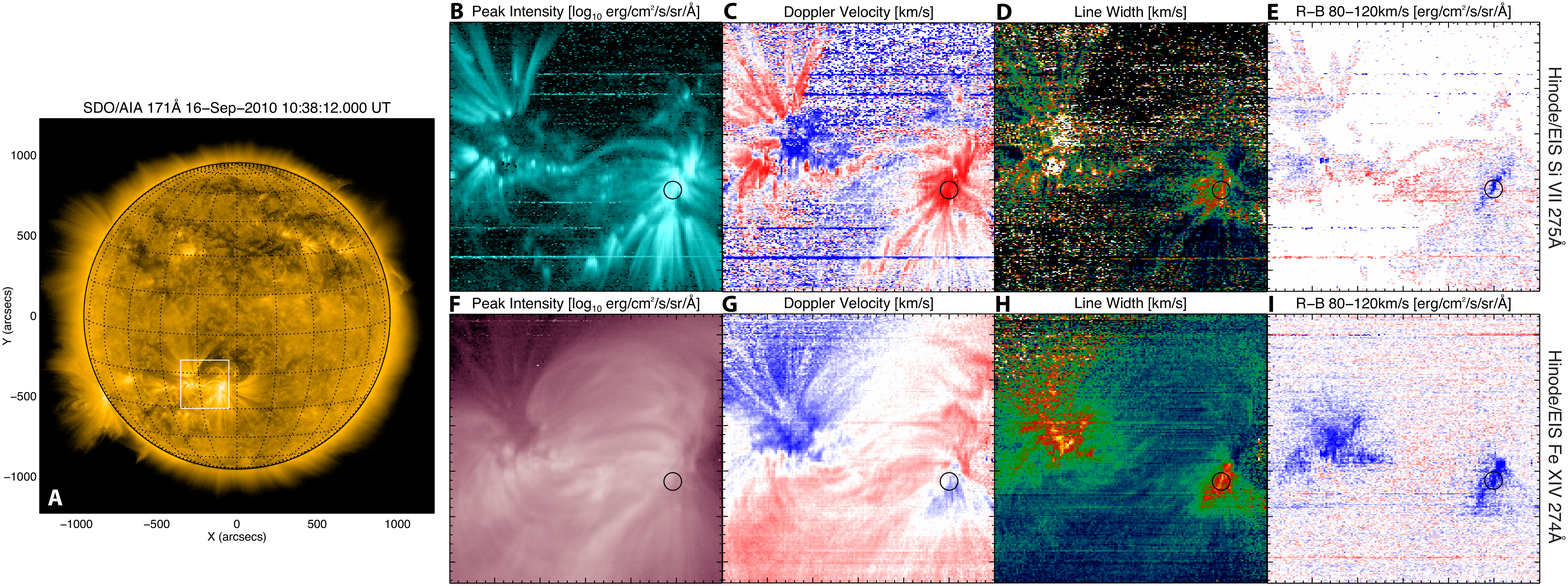}{fig1}{Contextual imaging and diagnostics 
of the observed active region. {\em (A)} SDO/AIA 171 \AA\/ image taken
at 10:38:12 UT on September 16, 2010. The white box indicates the
observable area analyzed in {\em Hinode}/EIS data. {\em (B) and (F)}
Intensity images from {\em Hinode}/EIS. The top row displays 
\ion{Si}{vii}~275~\AA\/ and the bottom shows \ion{Fe}{xiv}~274 \AA. 
{\em (C) and (G)} Doppler Velocities for the active region. {\em (D)
and (H)} Line profile widths. {\em (E) and (I)} RB measurements for
each profile, integrated from 80 to 120~km~s$^{-1}$. The circle
indicates the fan under consideration for our detailed analysis, where
blue asymmetry is clearly visible.}

\section{Data and Analysis}
The observations we study were taken by the EIS on 10:38:14 UT,
September 16, 2010, of NOAA AR 11106. This EIS study (435) was
supported as part of HOP 159 by SDO (taking its standard data products
at 12~s cadence) and allows us to obtain deep exposures (60~s) of the
emission in windows around the \ion{Si}{vii}~275 \AA, \ion{Si}{x}~258~\AA\/
and 261~\AA, \ion{Fe}{xii}~195~\AA, \ion{Fe}{xiii}~202~\AA,
\ion{Fe}{xiv}~264~\AA\/ and 274~\AA\/ lines---covering equilibrium formation
temperatures from 0.7~MK to 2~MK. The SDO data were recovered from the
SDO/JSOC at level 1 having being fully calibrated and corrected. The
EIS spectra were prepared using the standard IDL {\em SolarSoft} routines
distributed by the instrument team. For the purpose of illustration,
Fig.~\ref{fig1} shows the field of view of EIS (white rectangle)
relative to that of SDO/AIA in panel A.

Single Gaussian fit analysis was performed to each of the EIS spectral
windows in turn---the results from the extrema of the temperature
range are shown in panels B-D for \ion{Si}{vii}~275~\AA\/ and F-H for
\ion{Fe}{xiv}~274~\AA{}. Clearly, there is a stark difference in the
emission structure at these temperatures. In addition, we perform an
RB line profile asymmetry analysis \citep[][]{2009ApJ...701L...1D} to
the data, the results of which are shown in panels E and I integrated
from 80--120~km~s$^{-1}$. Briefly, the RB analysis entails fitting a
Gaussian {\em on a linear model background} to the line profile to
determine line center; interpolating the spectra to 10 times their
native resolution; measuring the emission in a narrow (20~km~s$^{-1}$)
range of velocities stepping symmetrically away from the line center
once the gradient in the background is removed; the RB measure at the
selected velocity is determined as the difference between the emission
in the red and blue wings. We note from Fig.~\ref{fig1}I that the
asymmetries present in the \ion{Si}{vii}~275~\AA\/ line of this region
are weak (see below).

\articlefigure{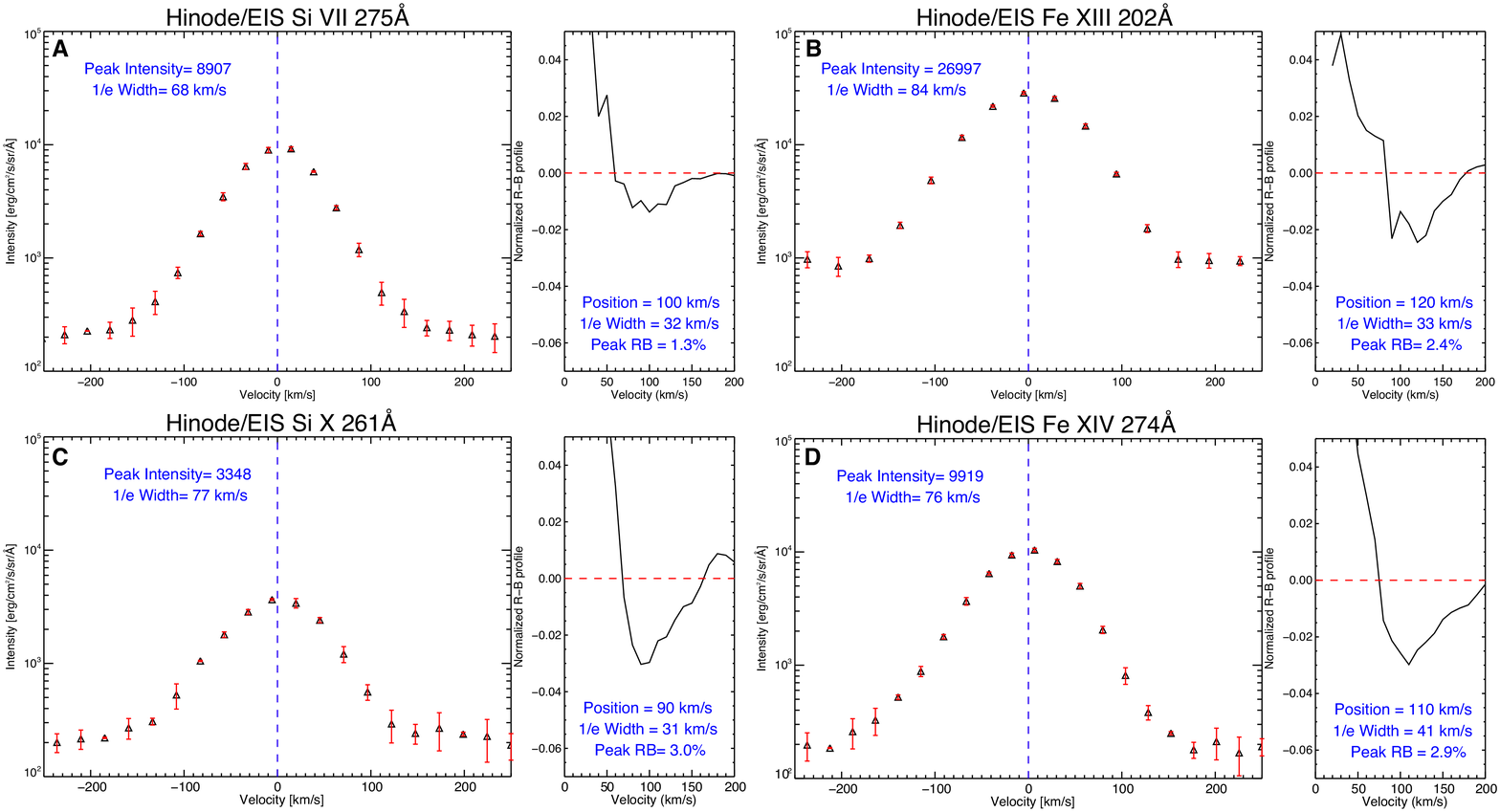}{fig2}{In pairs these panels show a 
sample of the EIS line profiles ({\em left}) and RB analysis of that
profile ({\em right}) in the region indicated in Fig.~\ref{fig1} that
sample the temperature space. The profiles are fitted with a single
Gaussian (and a linear background)---this is needed to determine the
centroid of the line that is shown as a blue dashed vertical line. The
profiles shown have had the background gradient removed to emphasize
the presence of the residual asymmetry. The right panel in each pair
shows the peak velocity of the RB second component, the 1/e width of
the RB second component, and approximate magnitude of the RB maxima.}

After identifying a common region of blue wing asymmetry (the small
circled region in Fig.~\ref{fig1}) we form an average line profiles
(over a $3 \times 3$ pixel range) to illustrate behavior over
temperature\footnote{The purpose of taking a spatial average for this
study is to get an accurate measure of these RB asymmetry properties
although we acknowledge that they can vary from one pixel to its
neighbor.}. Those average (background-gradient-removed) line profiles
are presented in the left-hand portions of Fig.~\ref{fig2} along with
the errors determined in the reduction process (red vertical bars) and
the location of the single Gaussian line center fit.  For these
average profiles (six in total, we show four---one covering each
temperature range) we perform the RB analysis and show the results in
the right-hand portions of the panels along with the determined
properties of the blue second component detected: its central
velocity, 1/e width, and magnitude normalized to the single fit peak
intensity. We note that in the majority of the analyzed spectra, the
background is not a constant across wavelength, but is often tilted
(sometimes severely in \ion{Si}{vii}~275~\AA). We account for this in our
analysis by using the linear background fit to estimate the background
contribution to the RB measure (in the case of a constant background
the RB contribution will be identically zero) such that the real
background is subtracted. For the average line profiles there are
visible asymmetries in the blue wing---the RB analysis is a means of
quantifying the their properties. In all cases, at this location, we
see signatures of a second component in the wing that extend from 80
to 150~km~s$^{-1}$, peaking around 100~km~s$^{-1}$---consistent with
previously published values \citep[e.g.,][]{2009ApJ...701L...1D}.

\articlefigure{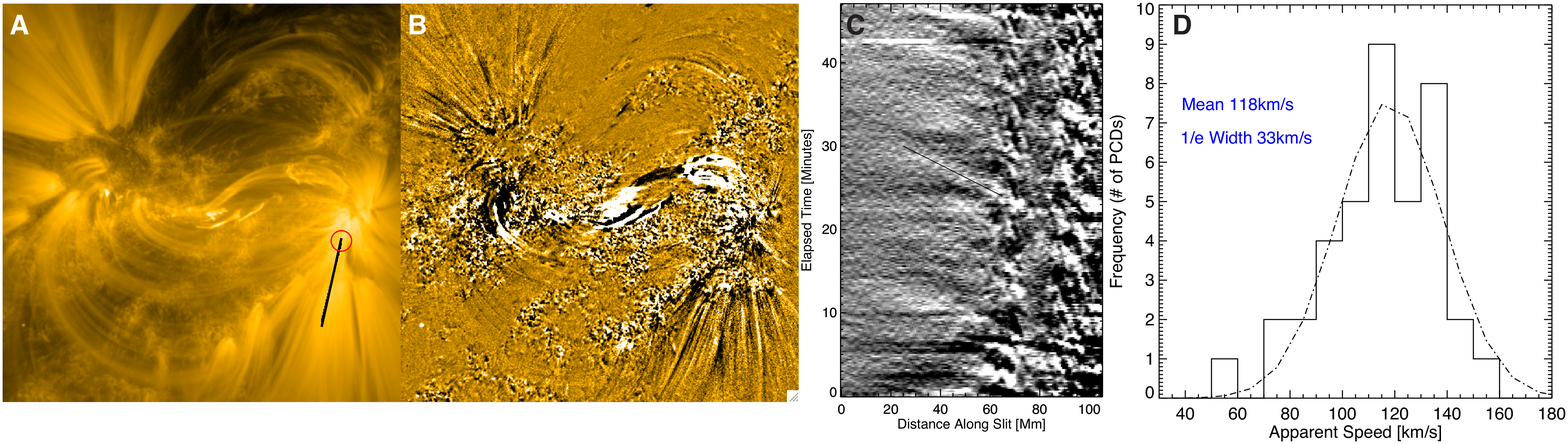}{fig3}{Studying the propagating coronal 
disturbances (PCDs) in the fan region. {\em From left to right} we
show sample AIA 171 \AA{} image, and running difference images, where
the former has the location of the ``slit'' on the fan to study the
PCD velocity distribution, the space-time plot determined from the
slit location, and the distribution of PCD apparent speeds from
sampling five slit angles rooted at the same point. The interested
reader can see a movie of panels A and B in an online movie---the link
is provided in the text.}

One of the goals of this paper is to establish the width of the second
component---that causing the blue wing asymmetry. Is it relatively
narrow (approximately the same width of the core emission) as the RB
analysis will have you believe or is it broad (significantly greater
than the core emission) as determined in \citet{2010A&A...521A..51P}?
We make the assumption that the second component is not related to the
thermal width of the emission and has a strong effect on the
non-thermal width---a connection that is visually supported by the RB
and line width panels of Figure~\ref{fig1}. Also, double Gaussian fits
have a very non-linear parameter space \citep[][]{1998A&AS..132..145M}
and require a thorough investigation when the sampling of the spectral
line is relatively poor---as it is with EIS---so we leave any
evaluation that the RB analysis will permit a more strongly
constrained double Gaussian fit that is minimally consistent with the
data for a future exercise. Instead we look for other avenues of data
support in understanding the second emission component. To this end we
use the high resolution observations of SDO/AIA to study the
distribution of apparent (plane-of-the-sky) speeds seen in the
propagating coronal disturbances (PCDs) rooted in the same fan
region\footnote{See the online supporting movie
\url{http://download.hao.ucar.edu/pub/mscott/Hinode4/f3.mov}}. 
\citet{2009ApJ...706L..80M} have already demonstrated the common 
locations and speeds of blue-wing asymmetries
and PCDs, here we investigate the PCDs seen to propagate at a small
range of angles in the fan and determine the likely distribution of
apparent speeds, and hence a possible ``width'' of the second
blue-shifted component. Forming a 100~Mm long slit inclined at an angle
of 25\deg\/ (see Fig.~\ref{fig3}) we use space-time plots to
characterize the motion of the PCDs as shown in panel C of
Fig.~\ref{fig3} and determining the speed using a linear fit to the
diagonal intensity signature of the PCD. The distribution of PCD
speeds shown in panel D was determined by sampling five space-time
plots from slits varying by $\pm 2\deg$ and rooted at the same
location in the fan. We see that the mean apparent speed of the PCDs
is 118~km~s$^{-1}$ with a 1/e width of 33~km~s$^{-1}$.

\articlefigure{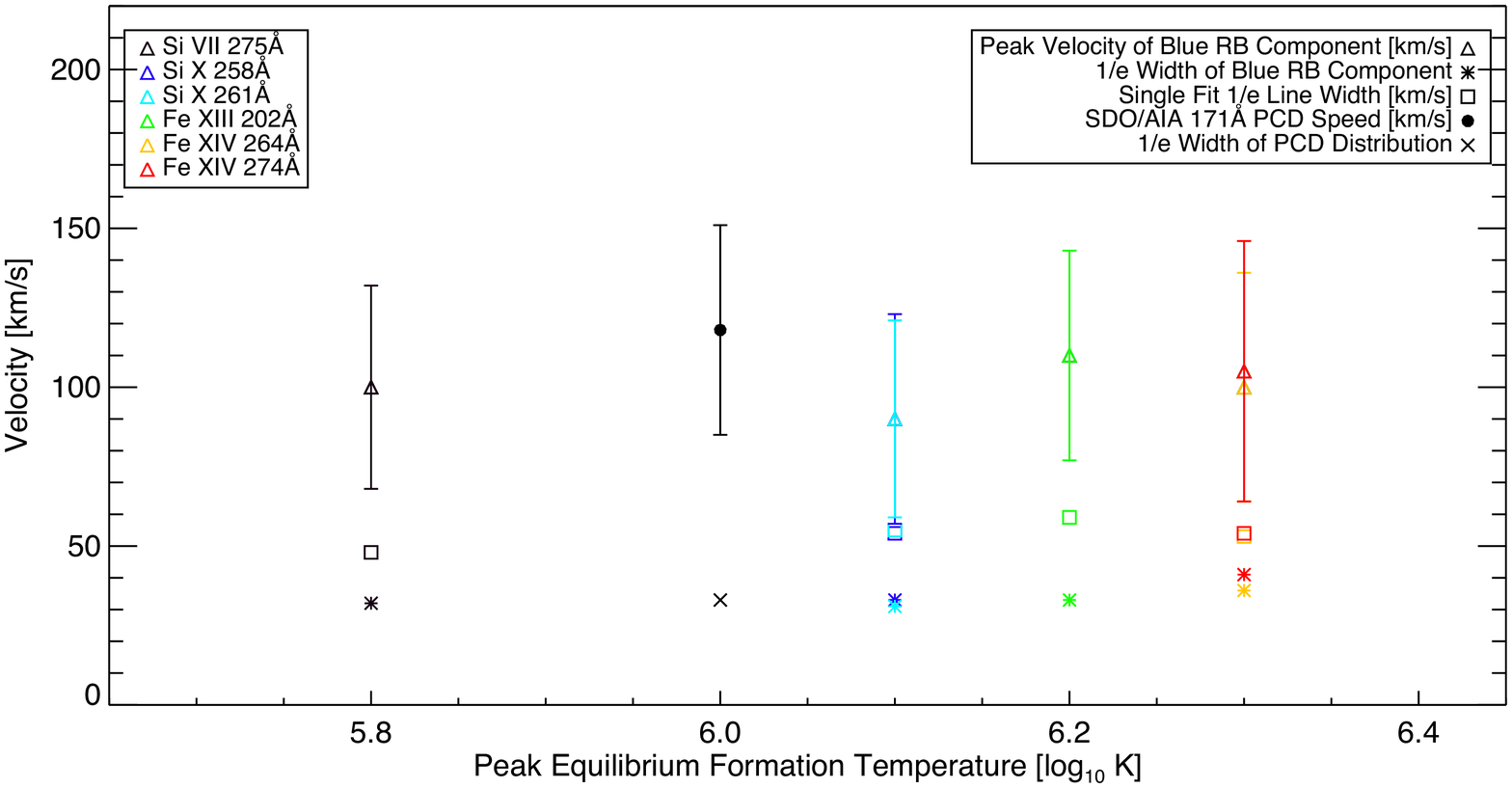}{fig4}{Information from the line 
profile and space-time plot analysis of the fan region.}

The information from the mean EIS line profiles of the fan root and
the study of the PCDs is collated in Figure~\ref{fig4}. The triangles
show the center position of the blue wing asymmetry with a 1/e width
demonstrated as a vertical error bar and pictorially as an
asterisk. The latter are for comparison with the single Gaussian fit
1/e width determinations for each line (square symbols). In addition,
the PCD values of mean velocity and 1/e width are show for the AIA
171~\AA{} channel as a dot, vertical error-bar, and cross
respectively. Summarizing the plot we see little evidence that there
is a strong temperature dependence of the RB asymmetry position or
width, the former being largely consistent with the AIA determined
speeds, with the obvious provision of directly comparing line-of-sight
(LOS) and plane-of-sky (POS) speeds. Furthermore, the widths of the RB
measure second component ($\sim$35~km~s$^{-1}$) are systematically
smaller than the 1/e width of the single Gaussian fits
($\sim$75~km~s$^{-1}$) and are commensurate with the width of the PCD
speed distribution.

\section{Discussion}
Based on the evidence presented we can comment on the minimally
consistent picture of asymmetries in EIS line profiles observed in
portions of an active region: 
\begin{enumerate}
\item There is (more often than not) more
than one (non-blend) component of emission in an EIS resolution
element; consider a $\chi^{2}$ map of a single Gaussian fit---does the
image contain structure that resembles the peak intensity pattern? 
\item There is a preponderance of asymmetry in the blue wing of the emission
lines of at the magnetic footpoints of non-flaring active
regions---based on the RB analysis of the data considered, that
asymmetry is centered at a similar speed, around
$\sim$110~km~s$^{-1}$.
\item The background of the line profiles must be
carefully considered and accounted for when quantifying any profile
asymmetry. 
\item The width of that component giving rise to the blue wing
asymmetry is of order 40~km~s$^{-1}$ while the line widths are of
order 70~km~s$^{-1}$, they are not considerably larger.
\item The
properties of the RB analysis determined component is consistent with
the properties of the PCDs clearly observed in SDO/AIA image
sequences.
\end{enumerate}

These points (the last couple in particular) appear to indicate that
the asymmetric line profiles observed by EIS are not always consistent
with a very broad weakly shifted ``pedestal'' component and an
unshifted core as presented by \citet{2010A&A...521A..51P}
\citep[and][]{2001A&A...374.1108P}. This comes, of course, with the
provision that it is very difficult to quantify the degree of
Alfv\'{e}nic transverse motion of the PCDs that can symmetrically
broaden this component---a 20~km~s$^{-1}$ PCD Alfv\'{e}nic motion
\citep[][]{2008ApJ...673L.219M,2011Natur.475..477M} would widen that
component from 40 to 45~km~s$^{-1}$ (using a quadratic summation) and
not by considerably more. The consistency of the RB peak speed
supports recent observations
\citep[][]{2010ApJ...722.1013D,2011ApJ...727L..37T} which counter the
notion that all PCDs are slow-mode longitudinal waves
\citep[e.g.,][]{2009ApJ...696.1448W,2010ApJ...724L.194V}. Further, we
note that the asymmetries can be small, very small, and so background
emission can mask them or enhance them depending on the direction of
the slope---these factors may have played into the blanket
determination of \citet{2010ApJ...715.1012B} that asymmetries are not
present in cooler lines. Despite the above points, we anticipate that
this location may be abnormal, the magnitude of the asymmetries is not
uniform from location to location and line to line (see, e.g., the RB
maps of Fig.~1), but very detailed analyses are required to
unequivocally determine the properties of asymmetries. The
asymmetries, as clearly highlighted in the innovative work of
\cite{2008ApJ...678L..67H}, have a center-to-limb dependence that
would suggest that the viewing geometry is very important in
determining if, when, where, and how much of an asymmetry is observed
and its physical origins \citep{2011ApJ...732...84M}. There is also
the issue of the apparent red-shift in the single Gaussian fit to the
cooler lines (cf. panels C and G of Fig.~\ref{fig1}) in places where
blue-wing asymmetries exist---an indication that the same magnetic
structure has cooling (radiatively losing energy as it falls) and
heated upflowing material contributing to the emission. As illustrated
by \citet{2009ApJ...701L...1D}, this is a complicating factor to many
conceptual models of the corona \citep[e.g.,][]{2011ApJ...727...58W}
and one that will be pursued in due course.

SDO/AIA image sequences provide invaluable contextual information to
help diagnose the origins of line profile asymmetries. Because the
PCDs are not shifted out of the broad imaging passbands their
propagation can be ``tracked'' along the coronal structure
\citep[e.g.,][]{2008SoPh..252..321M}. Using this motion tracking
algorithm and extrapolations of the photospheric vector magnetograms
we hope to infer the correct propagation angle of the PCDs relative to
the line of sight. This will allow us to unify the LOS and POS
velocity measures and validate the measurements presented. Therefore,
in future work we will explore the properties of the PCDs across the
same temperature space sampled by this EIS sequence, investigating
also the impact of the magnetic field orientation on the magnitude of
asymmetry observed. These asymmetries cannot be ignored: they may
carry invaluable information about the initial transport of mass into
the corona
\citep[][]{2011Sci...331...55D}. As such they {\em must} be
investigated fully before we begin to apply double Gaussian fitting
methods to spectroscopic data in an ad hoc fashion---a warning also
carried by \citet{2010A&A...521A..51P}.

\section{Conclusion}
We have performed a detailed analysis of a small fan root using RB
analysis of {\em Hinode}/EIS spectra and SDO/AIA imaging. We see that,
at this location, asymmetries are present in the blue wing of the
emission lines observed that are minimally consistent with a picture
where there are two components of emission present. The properties of
those asymmetries indicate that the velocities of the second emission
component are (relatively) consistent across temperature, they are
also of similar magnitude and width. Further, the RB-determined
asymmetry properties are also consistent with the distribution of
apparent speeds measured from the propagating coronal disturbances
visible in SDO/AIA image sequences originating at the same
location. Based on these points and the limited region studied, we
conclude that there is no evidence from our preliminary analysis that
this second emission component at the selected location is broader
than the main component of the line---such investigations hold vital
clues to the origin and mechanism by which material is ``pumped'' into
the corona.

\acknowledgements The effort was supported by external funds 
(NNX08AL22G, NNX08AU30G from NASA and ATM-0925177 from the US
NSF). NCAR is sponsored by the National Science Foundation.

\bibliography{hinode4}
\end{document}